\documentclass[runningheads]{svmult}

\usepackage{makeidx}   
\usepackage{graphicx}  
\usepackage{subeqnar}  
\usepackage{multicol}  
\usepackage{physprbb}  
\makeindex             

\newcommand{\avejankah}[1]{{\langle#1 \rangle}}
\newcommand{\rundjankah}[1]{{\left( #1 \right)}}
\newcommand{\lajankah}{\mathrel{\vcenter
     {\offinterlineskip \hbox{$<$}\hbox{$\sim$}}}}
\newcommand{\gajankah}{\mathrel{\vcenter
     {\offinterlineskip \hbox{$>$}\hbox{$\sim$}}}}
\begin{document}
\title*{Supermassive Stars: Fact or Fiction?}
%
%
\toctitle{Supermassive Stars}
%
%
\titlerunning{Supermassive Stars}
%
\author{Hans-Thomas Janka
}
\authorrunning{Hans-Thomas Janka}
%
%
\institute{Max-Planck-Institut f\"ur Astrophysik,
           Karl-Schwarzschild-Str.~1,
           D-85741 Garching, Germany}

\maketitle              

\begin{abstract}
Supermassive black holes are now realized to exist in the
centers of most galaxies. The recent discoveries of luminous 
quasars at redshifts higher than 6 require that these black 
holes were assembled already when the Universe was less 
than a billion years old. They might originate from the 
collapse of supermassive stars, a scenario which could ensure
a sufficiently rapid formation. Supermassive stars are 
dominated by photon pressure and radiate at their Eddington 
limit, which drives their quasi-static evolution to a final
relativistic instability. Above some critical value of the 
metallicity, their collapse can lead to a gigantic explosion,
powered by the energy release due to hydrogen burning, but
below this critical metallicity their collapse inevitably 
ends in the formation of a black hole, accompanied by the
emission of huge amounts of energy in the form of neutrinos.
Although collapsing supermassive stars are the most powerful 
known burst sources of neutrinos, the associated conditions 
do not appear favorable for producing highly relativistic
outflows that can explain cosmic gamma-ray bursts.
\end{abstract}

\section{Introduction}

The existence of supermassive black holes (SMBHs) in 
most galaxies is now becoming a generally accepted 
fact~\cite{Ree98}. 
Increasingly better resolved observations reveal
large Doppler shifts of spectral lines from hot gas swirling  
around the galactic centers with huge velocities, indicating
extraordinarily high mass concentrations in remarkably
small volumes (Fig.~\ref{jankahf1}). The rapid orbital motions
of the stars in the cluster surrounding Sgr~A$^\ast$ 
in the Milky Way require 
the stabilizing gravitational attraction of a dark object with 
a mass of about $3\times 10^6\,M_{\odot}$~\cite{Eck97}. 
This, though, is near
the lower end of the empirical mass distribution
of supermassive galactic black holes. The black hole masses,
the largest of which exceed $10^9\,M_{\odot}$,
correlate with the luminosity of the elliptical-galaxy-like 
bulge part of the host galaxy and 
with the average line-of-sight random velocity
(``velocity dispersion'') of the stars in the host galaxy.
This indictes a correlation between black hole mass and 
galaxy mass and is interpreted as a hint to a direct connection 
between galaxy formation and black hole fueling~\cite{Kor95,Pag01}. 

SMBHs with masses of a million to a few billion 
solar masses are believed to be the engines that power active 
galactic nuclei ranging from faint, compact radio sources to
quasars that are brighter than the whole galaxy in which they live. 
The detection of quasars with redshifts larger than 6 requires
that these objects were formed only several hundred million years
after the Big Bang. The growth of black holes by accretion is
exponential, $M_{\mathrm{BH}} = M_{{\mathrm{BH}},0}\exp({t/\tau})$,
with a timescale $\tau \sim 4\times 10^7\,(\epsilon/0.1)/\eta\,$
years, where $\epsilon = L/(\dot M_{\mathrm{BH}}c^2)$ is the
radiation efficiency of the accreting black hole and 
$\eta = L/L_{\mathrm{Edd}}$ is the ratio of the luminosity $L$ to
the Eddington luminosity. Therefore SMBHs with 
$10^7$--$10^9$ solar masses need at least 10--20 e-foldings 
to assemble by accretion on seed black holes of 
10$\sim$1000$\,M_{\odot}$, which
might be the compact remnants of an early generation of massive 
or very massive stars~\cite{Heg01}.
Provided enough gas were available in the surroundings to be 
swallowed by such a stellar mass
black hole, and the efficiency of the black hole to absorb the
gas flow were as high as assumed above, the time for its growth
by accretion seems marginally short enough to explain the 
existence of quasars in the early Universe.

\begin{figure}[!htbp]
\begin{center}
\vspace{5.0truecm}
\end{center}
\caption[]{A 3,700 light-year-diameter dust disk around the
300 million solar-mass black hole in the center of the
elliptical galaxy NGC 7052 as observed by the Hubble
Space Telescope. (Credits: Roeland P. van der Marel (STScI),
Frank C. van den Bosch (University of Washington), and NASA)}
\label{jankahf1}
\end{figure}

A consistent and satisfactory picture of the formation process
of quasar black holes has not been developed yet~\cite{Ree01}.
A variety of different routes were suggested (for a 
review, see Ref.~\cite{Ree84}), including scenarios based on gas 
hydrodynamics, stellar dynamics, or combinations of both. 
Primordial gas clouds could collapse directly to SMBHs 
when their fragmentation is inhibited by radiation pressure or
magnetic fields~\cite{Loe94,Gne01}. 
Alternatively, when angular momentum plays a role and
the cooling timescale is much smaller than the viscous 
timescale, they might contract to form a supermassive 
disk~\cite{Wag69,Eis95}. 
If instead fragmentation of a primordial gas cloud occurred and a 
dense cluster of stars were born, stellar collisions and mergers 
might lead to a runaway growth of intermediate-mass black holes
(e.g., Refs.~\cite{Beg78,Qui90,Ebi01,Por02}).
A dense cluster composed of neutron
stars or black holes as the compact remnants of massive stars
might be driven by the secular ``gravothermal catastrophe''
to the point of a final relativistic instability that happens
on a dynamical timescale~\cite{Zel65,Sha85} (see, however, 
Ref.~\cite{Beg78} for arguments against this picture). 

Some of the proposed scenarios are envisioned 
to lead to the build-up of a supermassive star
(SMS) as an intermediate stage of the evolution before the final 
gravitational instability sets in and the collapse to a SMBH takes
place. Further growth of a seed black hole, possibly a SMBH, by 
accretion, could be linked to processes during the formation or 
evolution of the bulge of a galaxy (e.g., 
Refs.~\cite{Sch99,Ume01}). Galaxy formation or
interaction could directly result in the black hole feeding
that makes quasars shine, and bigger galaxies might be able to
provide more fuel. This might explain why more massive
galaxies contain more massive black holes~\cite{Kor95}.

\section{Supermassive Stars: Some Basic Facts}

Supermassive stars (SMSs) are equilibrium configurations that
are dominated by radiation pressure. Their temperature is low
enough that electron-positron pairs do not play a role. 
Baryons yield only a minor contribution to the equation of state.
At some point of their evolution SMSs
collapse due to a general relativistic gravitational 
instability~\cite{Hoy63,Ibe63,Cha64,Fow64,Fow66,Zel67,Bis67,Sha83}.

SMSs can have masses between $\sim$10$^4\,M_{\odot}$ and about
$10^8\,M_{\odot}$. Since they are expected to be fully 
convective~\cite{Sha83} (a formal argument can be found in
Ref.~\cite{Loe94}), they are isentropic and their structure
can be well described by a Newtonian polytrope with $n = 3$ or 
$\gamma = 1+1/n = 4/3$. With their specific entropy being nearly
constant, the adiabatic polytropic constant $\gamma$ is roughly
equal to the local adiabatic index $\Gamma = [{\mathrm{d}}(\ln P)/
{\mathrm{d}}(\ln \rho)]_s$. To good accuracy
the entropy per nucleon in a SMS, $s_{\mathrm{SMS}}$,
is given by the radiation entropy (in units of Boltzmann's 
constant $k$)
\begin{equation}
{s_{\mathrm{SMS}}\over k}\,\approx\,{s_{\mathrm{r}}\over k}\,=\, 
(1.22\times 10^{-22})\,{T^3\over \rho}\,=\,
0.94\,\rundjankah{{M\over M_{\odot}}}^{\! 1/2}\,,
\label{jankaheqs}
\end{equation}
where $M$ is the mass of the star, $\rho$ the density in 
g$\,$cm$^{-3}$, and $T$ the temperature in K 
(see, e.g., Refs.~\cite{Sha83,Ful86}).
SMSs with large masses have entropies much higher than typical
astrophysical plasmas. This suggests that dissipative processes
(shocks, cloud-cloud collisions, turbulence) must be invoked
for generating appropriate conditions for the formation of
such supermassive equilibrium configurations.

The evolution of SMSs proceeds on a Kelvin-Helmholtz timescale 
and is driven by the loss of energy and entropy through radiation 
and --- in case of rotation being important --- the loss
of angular momentum, e.g. via mass shedding. Because the 
pressure is dominated by radiation, the luminosity of SMSs
is close to the Eddington limit,
\begin{equation}
L_{\mathrm{SMS}}\,=\,L_{\mathrm{Edd}}\,=\,
{4\pi GcMm_p\over \sigma_{\mathrm{T}}}\,=\,
1.3\times 10^{38}\,{M\over M_{\odot}}\ {\mathrm{ergs\,s}}^{-1}\,,
\label{jankaheqlum}
\end{equation}
with $\sigma_{\mathrm{T}}$ being the Thompson cross section and 
$m_p$ the proton mass. 

Although plasma corrections (due to nuclei and 
electrons\footnote{For reasons of simplicity, a pure hydrogen 
plasma is assumed here. The expressions for the general case 
can be found in Refs.~\cite{Ful86,Shi98}).}) 
and general relativistic effects 
are small, neither of both can be neglected for the
evolution of a configuration which is so close to the 
pathological, i.e., dynamically marginally stable, case of a 
$\gamma = 4/3$ polytrope (for a discussion, see Ref.~\cite{Sha83}).
Pressure contributions by plasma components
raise the adiabatic index of the equation of state, 
$\Gamma$, to a value above 4/3~\cite{Cha39}:
\begin{equation}
\Gamma_{\mathrm{SMS}}\,=\,1 + {P\over \varepsilon}\,\approx\,
{4\over 3} + {\beta \over 6}\,.
\label{jankaheqgam}
\end{equation}
Here $\varepsilon$ is the total internal energy density without
rest-mass energy, and the second expression is correct to first
order in the ratio of the gas pressure to the radiation pressure,
$\beta = P_{\mathrm{g}}/P_{\mathrm{r}} = 8/(s_{\mathrm{r}}/k) \ll 1$.

General relativity leads to the existence of a maximum 
for the equilibrium mass as a function of the central density for 
SMSs with constant entropy. This implies that general relativistic 
stars of mass $M$, which are supported both by radiation and gas 
pressure, have a maximum, i.e., ``critical'', central density:
\begin{equation}
\rho_{\mathrm{crit}}\,=\,2\times 10^{18}\,\rundjankah{{M\over
M_{\odot}}}^{\! -7/2}\ \mathrm{g\,cm}^{-3}\,. 
\label{jankaheqrhocrit}
\end{equation}
For higher central densities the nonlinear effects of gravity
have a destabilizing influence to radial perturbations. The
central temperature at the onset of the gravitational instability 
is 
\begin{equation}
T_{\mathrm{crit}}\,=\, 2.5\times 10^{13} \rundjankah{{M\over
M_{\odot}}}^{\! -1}\ \mathrm{K}\,,
\label{jankaheqtemp}
\end{equation}
and the corresponding equilibrium energy can be found to be
\begin{equation}
E_{\mathrm{crit}}\,=\,-3.6\times 10^{54}\ {\mathrm{ergs}}\,,
\end{equation}
which is independent of $M$.
The redshift factor at the surface of a supermassive star 
(with radius $R = R_{\mathrm{crit}}$) at this point of the 
evolution is
\begin{equation}
\rundjankah{{GM\over Rc^2}}_{\! \mathrm{crit}}
\,=\,{1\over 2}\,{R_{\mathrm{s}}\over R_{\mathrm{crit}}}
\,=\,0.63 \rundjankah{{M\over M_{\odot}}}^{\! -1/2}\,, 
\label{jankaheqrs}
\end{equation}
which is in the range $10^{-2}$....$10^{-4}$ for 
$M = 10^4\,M_{\odot}$....$10^6\,M_{\odot}$ and thus indeed
very small so that the configuration is much larger than its 
Schwarzschild radius,
$R_{\mathrm{crit}}/R_{\mathrm{s}} = 0.794\sqrt{M/M_{\odot}}$.
Nevertheless, general relativity plays a crucial role for the 
evolution.

The gravitational instability sets in when the effective adiabatic
index of the configuration (Eq.~\ref{jankaheqgam}) drops below a 
critical value, which is approximately given by
\begin{equation}
\Gamma\,<\,\Gamma_{\mathrm{crit}}\,\approx\,
{2\over 3}\,{2-5\eta\over 1-2\eta}\,+\, 1.12\,
{R_{\mathrm{s}}\over R} \,.
\label{jankaheqcrit}
\end{equation}
$R_{\mathrm{s}} = 2GM/c^2$ is the Schwarzschild radius of
the star and $\eta = T/|W|$ is the ratio of the rotational
energy to the gravitational potential energy. For $\eta = 0$, 
Eq.~(\ref{jankaheqcrit}) becomes 
$\Gamma_{\mathrm{crit}}=4/3+1.12R_{\mathrm{s}}/R$
(see Ref.~\cite{Mis73} and references therein).
Therefore the gravitational collapse begins when the 
plasma contribution to the equation of state does not raise the 
adiabatic index sufficiently above $4/3$ to compensate for the 
destabilizing influence of general relativity. The latter grows
as the radius of the star shrinks during its evolution. Rotation
has a stabilizing effect and can hold up the collapse if
\begin{equation}
\eta\,>\,\eta_{\mathrm{crit}}\,\approx\,{1\over 2}\,
{4-3(\Gamma - 1.12 R_{\mathrm{s}}/R)
\over 5-3(\Gamma - 1.12 R_{\mathrm{s}}/R)} \,.
\end{equation}
If $\eta > 0$ remains a constant during the equilibrium evolution, 
the final instability is again reached when the relativistic
terms become too large.

Since the central density and temperature are higher for stars
with smaller masses, and both increase during the quasistatic 
evolution, the creation of electron-positron pairs can be neglected
for $M\gajankah 10^4\,M_{\odot}$ (i.e., $T \le T_{\mathrm{crit}} 
\lajankah 2.5\times 10^9\,$K; Eq.~\ref{jankaheqtemp})~\cite{Bon84}.
A comparison of the nuclear energy generation rate with the photon
luminosity (Eq.~\ref{jankaheqlum}) shows that nuclear burning is
also irrelevant prior to the gravitational instability for SMSs
with masses in excess of a few $10^5\,M_{\odot}$~\cite{Fow66,Zel67}.

While radiating energy at its Eddington limit, the SMS evolves 
on a timescale 
\begin{equation}
t_{\mathrm{KH}}\,=\,{|E_{\mathrm{crit}}|\over L_{\mathrm{SMS}}}
\,=\,2.8\times 10^{16}\rundjankah{{M\over M_{\odot}}}^{\! -1}\,
\mathrm{s}
\label{jankaheqkh}
\end{equation}
with essentially constant mass but decreasing entropy and energy
until the critical configuration for the general relativistic  
instability is reached and the catastrophic collapse sets in.
Only when the thermal (Kelvin-Helmholtz) timescale is longer
than the hydrodynamical timescale,
\begin{equation}
t_{\mathrm{hydro}}\,\sim\,(G\rho)^{-1/2}\,=\,
2.7\times 10^{-6} \rundjankah{{M\over M_{\odot}}}^{\! 7/4}\,
\mathrm{s}\,,
\end{equation}
there is an equilibrium phase of the evolution of the supermassive 
object. The two timescales become equal (about 10 years)
for $M\sim 10^8\,M_{\odot}$. Above this mass no hydrostatic
phase is possible. More typical SMSs with 
$M \sim 10^6\,M_{\odot}$ have a lifetime of about 1000 years.

Since SMSs are very close to the edge of instability,  
rotation can appreciably stretch their equilibrium
evolution. Considering the secular evolution of a
uniformly rotating configuration along the mass-shedding   
sequence,
Baumgarte and Shapiro~\cite{Bau99a} found a lifetime independent
of the stellar mass, $t \approx 9\times 10^{11}\,s$.
Moreover, they showed that the key nondimensional ratios
$R/R_{\mathrm{s}}$, $T/|W|$, and $Jc/(GM^2)$ ($J$ is the 
angular momentum) for a maximally and rigidly rotating 
$n = 3$ polytrope at the onset of radial instability are 
universal numbers, independent
of the mass, spin, radius, or history of the star:
\begin{equation}
\rundjankah{{T\over |W|}}_{\mathrm{crit}}\approx\, 0.009\,,\ \ 
\rundjankah{{Jc \over GM^2}}_{\mathrm{crit}} \approx\, 0.97\,,\ \
\rundjankah{{R_{\mathrm{p}}\over R_{\mathrm{s}}}}_{\mathrm{crit}}
\approx\,214 \,,
\end{equation}
with the polar radius $R_{\mathrm{p}}$ being roughly $2/3$
of the equatorial radius $R_{\mathrm{e}}$. This deformation 
reduces the luminosity by about 36\% below the usual Eddington
luminosity of a corresponding nonrotating star~\cite{Bau99b}.
The effects of differential rotation, where mass loss during the 
quasi-stationary evolution could be ignored, 
were discussed by New and Shapiro~\cite{New01}.

\section{The Death of Supermassive Stars}

Driven by energy loss through radiation and angular momentum
loss due to mass shedding, SMSs contract slowly in a 
quasi-static manner and approach the point of dynamical
instability as a consequence of the increasing effects 
of general
relativity. The subsequent catastrophic collapse can lead to
the formation of a SMBH~\cite{App72a} or,
for sufficiently large initial metallicity, to
a violent explosion that is powered by the release of nuclear 
energy through hydrogen burning in the CNO 
cycle~\cite{Bis68,App72b,Fri73,Fri74,Ful86}.

These events are associated with the emission of gigantic
amounts of energy in neutrinos~\cite{Woo86}, 
a fact that has nourished 
speculations that SMSs collapsing to black holes might be 
the sources of cosmic gamma-ray bursts~\cite{Ful98}.
In the case of rotationally deformed configurations, which
might encounter a dynamical bar mode instability that 
triggers the growth of nonaxisymmetric bars, the generation of
long-wavelength gravitational waves can be expected. Such
gravitational waves could be detectable by future space-based 
laser interferometers like LISA~\cite{Bau99a,New01,Sai02}.
Moreover, SMSs that 
were disrupted by explosions might have contributed to the 
enrichment of the gas in the young Universe with elements
heavier than helium, in particular of comparatively rare
isotopes like $^{13}$C, $^{15}$N, $^{17}$O, 
and $^{22}$Ne~\cite{Aud73}, and in case of SMSs with high
metallicities of $^{26}$Al~\cite{Hil87}. 

\begin{figure}[!htbp]
\begin{center}
\includegraphics[width=.99\textwidth,clip]{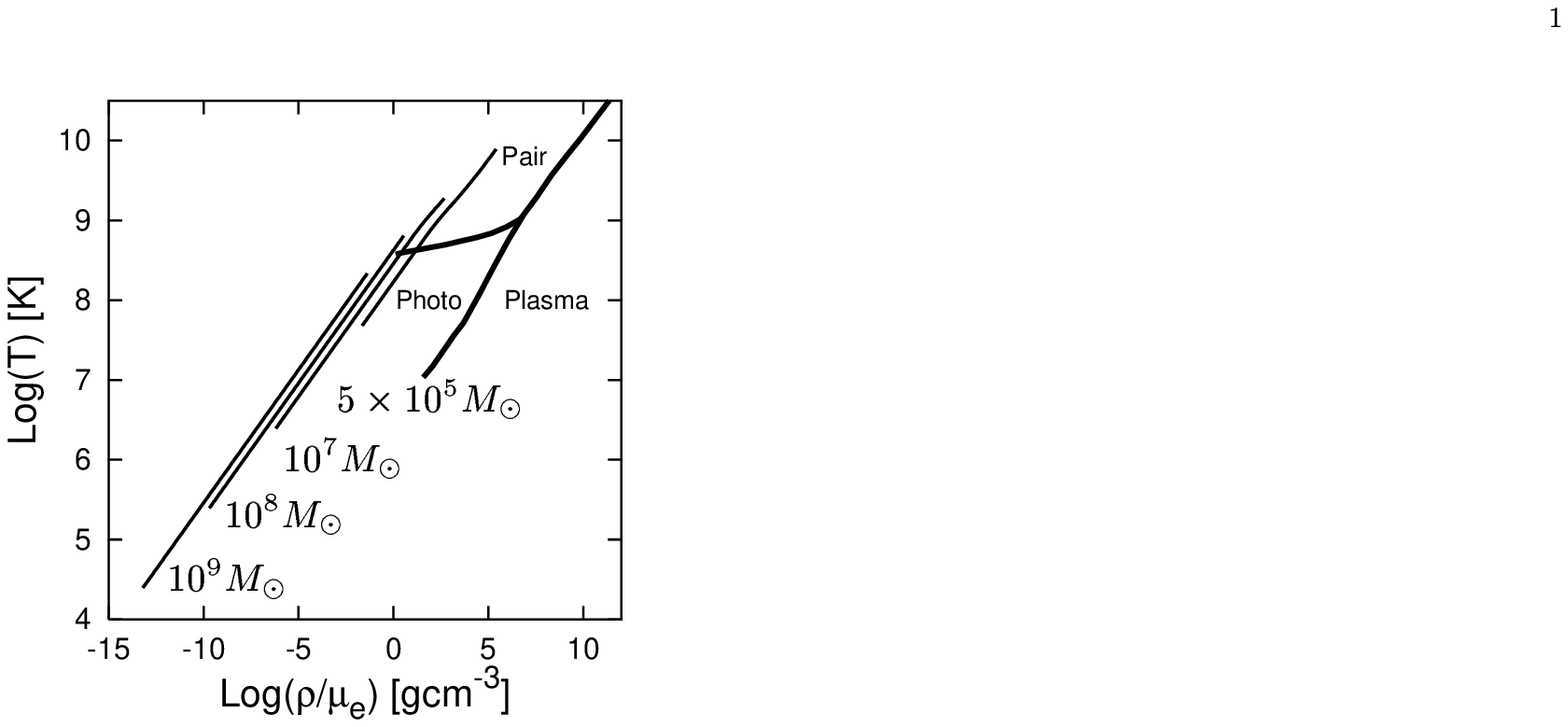}
\end{center}
\caption[]{Evolutionary tracks of the center of collapsing 
SMSs with different
masses in the $\rho$-$T$-plane ($\mu_{\mathrm{e}}$ is the mean
molecular weight). Initially the photo-neutrino production
yields the major contribution to the neutrino loss, during
the later phases of the collapse it is the pair annihilation
process. Plasmon neutrinos are negligible}
\label{jankahf2}
\end{figure}

The last investigation of nuclear burning during 
the collapse and explosion of SMSs was undertaken by Fuller 
et al.~\cite{Ful86}. They performed hydrodynamical calculations
with a post-Newtonian approximation to general relativistic
gravity, used a detailed equation of state including 
electron-position pairs, and took into account photon
and neutrino losses as well as all nuclear reactions 
for describing
hydrogen burning by the CNO cycle (limited by $\beta^+$ decays
of $^{14}$O and $^{15}$O) and by the rapid proton capture
(rp-) process that characterizes hydrogen burning at very
high temperatures ($T \gajankah 10^9\,$K; at such temperatures,
however, neutrino losses become dominant). Considering
nonrotating configurations, Fuller et al.~\cite{Ful86}
found that stars with a mass $M\gajankah 
10^5\,M_{\odot}$ and initial metallicities $Z \lajankah
0.005$ do not explode, whereas objects with $Z \gajankah
0.005$ (a value near the solar mass fraction of heavy 
elements!) do explode. The explosion energies range from
$2\times 10^{56}\,$ergs for stars of mass 
$M\approx 10^5\,M_{\odot}$ to $2.5\times 10^{57}\,$ergs
for $M\approx 10^6\,M_{\odot}$, and the photon luminosities
can exceed $10^{45}\,$ergs$\,$s$^{-1}$ for a period of
more than ten years. The nucleosynthesis in  
exploding SMSs is characterized by the production of a 
large amount of $^4$He and trace amounts of $^{15}$N and
$^7$Li. Deuterium production turned out to be negligible 
because this nucleus is too fragile to survive the high
temperatures during the explosion of supermassive objects
(see, however, Ref.~\cite{Ful97} for neutrino-induced
deuterium generation in the ejected envelope of exploding SMSs).
Since zero metallicity (nonrotating) SMSs do not blow up,
they, however, cannot be considered as a source of pre-galactic
helium. 

Recently Linke et al.~\cite{Lin00,Lin01} 
have performed simulations in
full general relativity of the collapse of nonrotating SMSs 
to black holes with the aim to calculate in detail the neutrino
(and antineutrino) emission of such events until the point 
when the emission is quickly terminated by the formation of
the event horizon. Their models also included electron-positron
pairs and plasma contributions besides photons
in the equation of state and were focused on cases where 
the energy release by nuclear burning is unimportant because
it is dwarfed by neutrino losses. 
In Fig.~\ref{jankahf2} the evolutionary tracks of the central
density and temperature of SMSs in the mass range between
$5\times 10^5\,M_{\odot}$ and $10^9\,M_{\odot}$ are plotted
on top of the regions of dominant energy loss by the 
neutrino-antineutrino pair production through the 
photo-neutrino process ($\gamma+e^\pm\to e^\pm+\nu+\bar\nu$),
electron-positron pair annihilation ($e^-+e^+\to\nu+\bar\nu$), 
and plasmon decay (${\widetilde{\gamma}} \to \nu+\bar\nu$). 

Energy losses by neutrino production become important only 
during the later stages of the collapse. Initially
the photo-neutrino process plays 
the dominant role, whereas shortly before the black hole 
forms, when most of the neutrino emission occurs, 
the temperature is so high that the pair-neutrino 
process takes over. Plasmon neutrinos yield a negligible 
contribution in all cases. 
The neutrino emission rates are extremely temperature dependent.
The energy production rate by electron-positron annihilation,
$Q_{\nu}$, rises like $T^9$ above the threshold temperature for 
pair formation and even more steeply ($Q_{\nu} \propto T^{20}$)
for temperatures just below $10^9\,$K when $e^+e^-$ pair
creation sets in~\cite{Ito96}. 

\begin{figure}[!htbp]
\begin{center}
\includegraphics[width=.49\textwidth, clip]{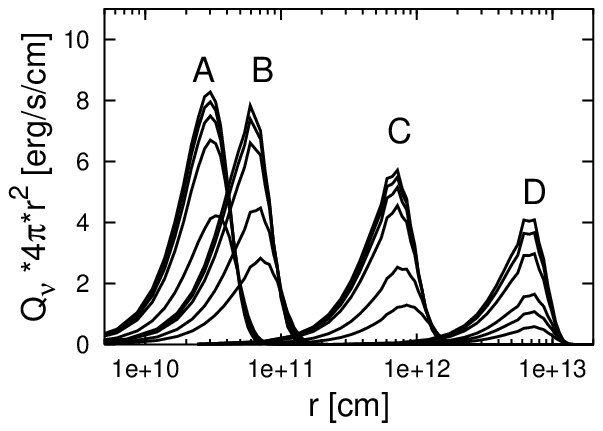}
\includegraphics[width=.49\textwidth, clip]{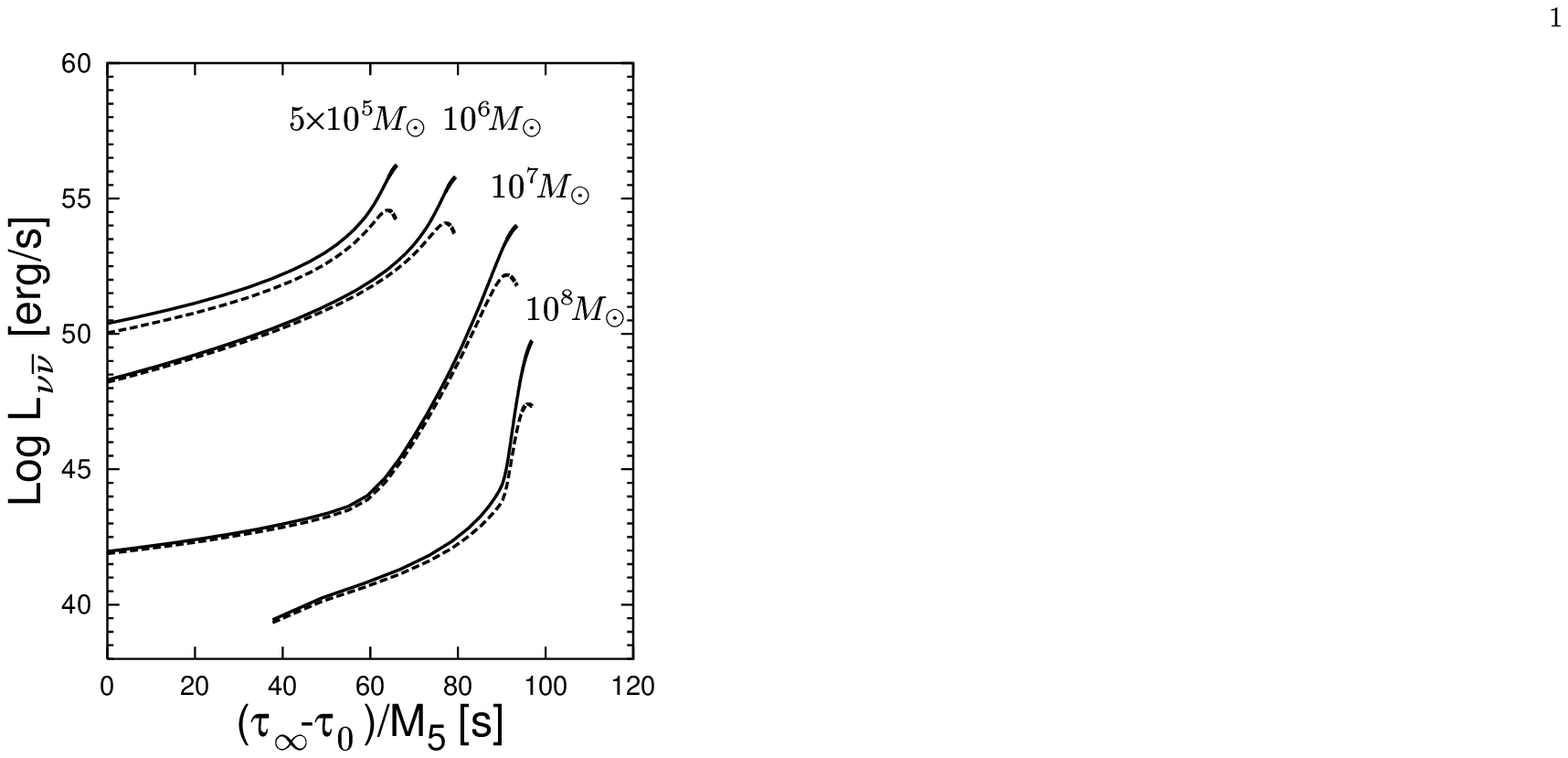}
\end{center}
\caption[]{
Radial profiles of the neutrino plus antineutrino emission
rate (times $4\pi r^2$; left) and luminosities of neutrinos
plus antineutrinos as a function of time (right) for
collapsing SMSs with different masses. The left plot gives the
quantity ${\mathrm{d}}L_{\nu\bar\nu}(r)/{\mathrm{d}}r$ for
$M = 5\times 10^5\,M_{\odot}$ (A), $10^6\,M_{\odot}$ (B),
$10^7\,M_{\odot}$ (C), and $10^8\,M_{\odot}$ (D) for
different epochs of the model evolution with the end of the
simulations being represented by the uppermost lines.
The corresponding scaling factors are $5\times 10^{44}$ (A),
$10^{44}$ (B), $2\times 10^{41}\,M_{\odot}$ (C), and
$2\times 10^{36}$ (D), respectively.
In the right figure the dashed lines include Doppler shifts and
general relativistic redshift effects, the solid lines do not.
The time is measured in
seconds with $\tau_\infty$ being the proper time for an observer
at infinity, shifted by the overall collapse timescale
($\tau_0 = 8\times 10^5\,$s, $1.7\times 10^6\,$s,
$8.0\times 10^7\,$s, and $3.2\times 10^9\,$s, respectively)
and scaled by $M_5 = M/(10^5\,M_{\odot})$}
\label{jankahf3}
\end{figure}

Although enormous amounts of energy are radiated away in neutrinos,
these energy losses are small compared to the internal energy 
or the gravitational potential energy of the star. The neutrino
losses are therefore essentially irrelevant in the energy budget
and the collapse can well be considered as adiabatic.
It proceeds nearly homologously so that the density
profile evolves in a self-similar manner. Deviations from this
ideal behaviour of a Newtonian $n = 3$ polytrope at a later
stage of the collapse are caused by the increasing influence
of general relativity and to a minor degree also by the possible 
formation of electron-positron pairs and the corresponding
reduction of the adiabatic index. Except for such effects 
that are associated with the equation of state or  
with the neutrino emission --- both of which 
are very sensitive to the maximum value of the temperature that is
reached during the implosion --- the collapse of SMSs of different
masses is found to be very similar. The stellar mass therefore 
simply acts as a scaling parameter, a fact which will be made use 
of in the following discussion.

\begin{figure}[!htbp]
\begin{center}
\includegraphics[width=.49\textwidth, clip]{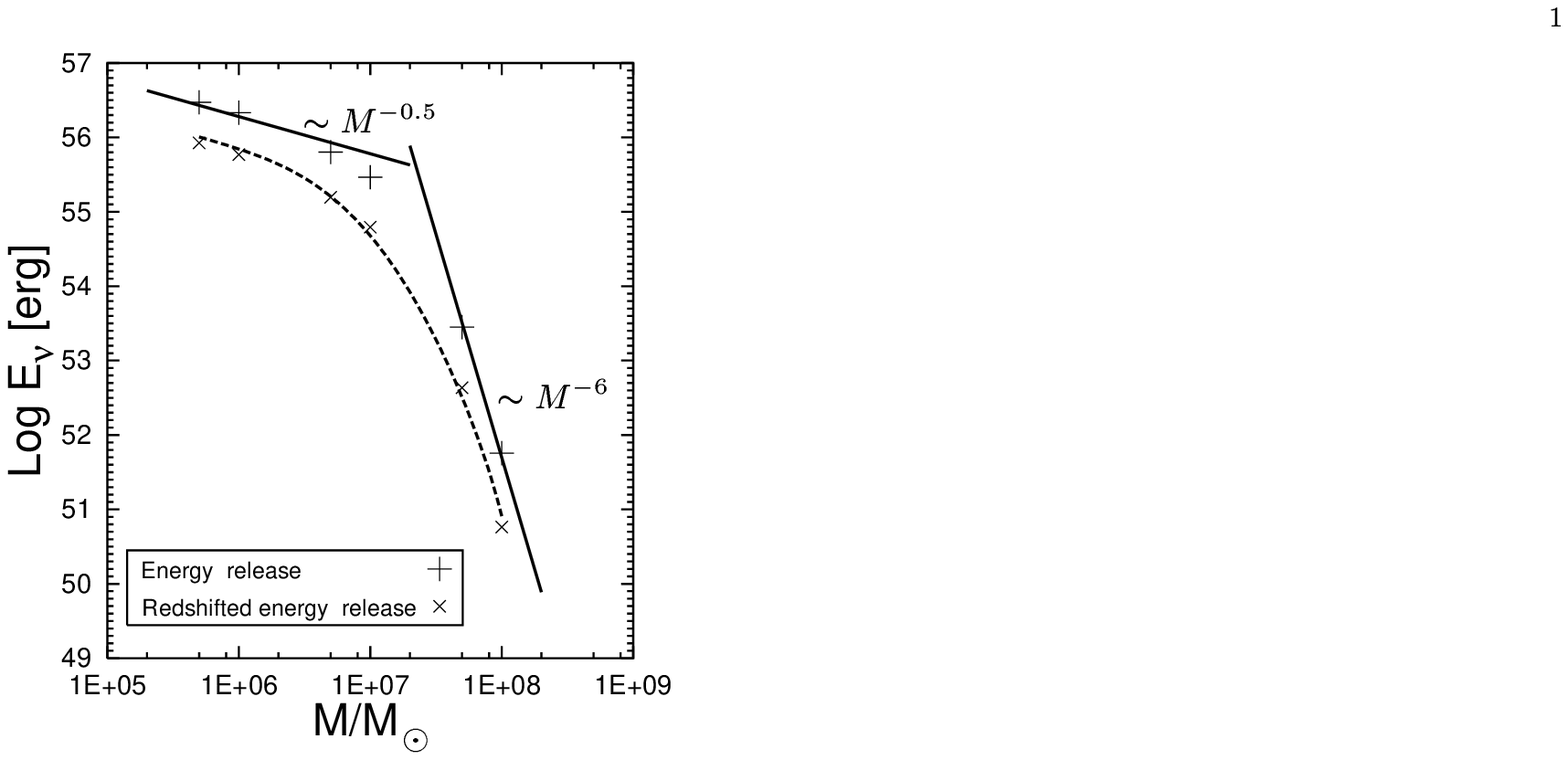}
\includegraphics[width=.49\textwidth, clip]{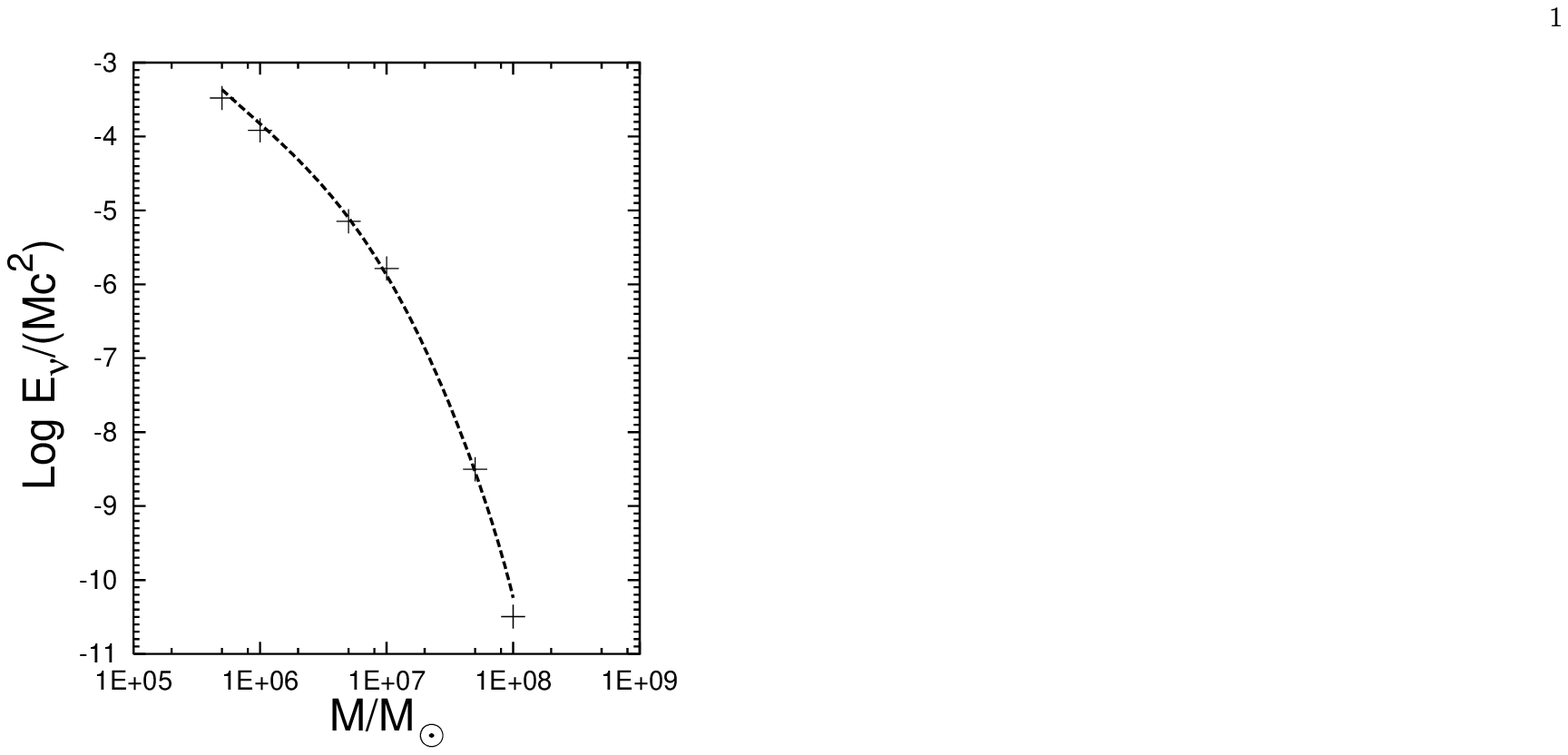}
\end{center}
\caption[]{Total energy release in form of neutrinos during
the collapse of SMSs to SMBHs (left). The mass of the star is given
on the abscissa. The dashed curve interpolates the computed
values (symbols) which include the effects of Doppler shift and
general relativistic redshift, while the solid lines interpolate
the results for the integral energies as measured by local
observers. The right plot shows the efficiency of the conversion
of rest-mass energy to neutrinos, $E_{\nu}/(Mc^2)$. In case of
less massive stars the higher core temperatures before black
hole formation imply much larger values for the total energy
emitted in neutrinos and for the conversion efficiency}
\label{jankahf4}
\end{figure}

The collapse timescale $t_{\mathrm{coll}}$ is roughly proportional 
to $R_{\mathrm{s}}/c \propto M$ and lasts between about 9 days
for a SMS with $5\times 10^5\,M_{\odot}$ and several years for
$10^7\,M_{\odot}$ stars. For more massive stars a meaningful
calculation of the duration of the phases of contraction and
implosion requires the inclusion of photon emission
(cf.\ Eq.~\ref{jankaheqkh}), which in fact was ignored in our
models. The collapse of the innermost $\sim 25$\% of the mass 
proceeds in a nearly coherent way with an approximately
homologous velocity profile. This part of the star therefore
forms a black hole first and determines the radius of the 
innermost apparent horizon, which is proportional to $M$.

Because the highest temperatures are reached at the end of the
collapse, the time interval of strongest neutrino emission is
also given by $t_{\mathrm{em}}\sim R_{\mathrm{s}}/c \propto M$. 
The peak of the neutrino production is located in a shell just
outside the innermost apparent horizon (Fig.~\ref{jankahf3}). 
The corresponding ``neutrino emission radius'' is therefore 
also proportional to the event horizon of the SMS, 
$R_{\nu}\propto R_{\mathrm{s}}\propto M$ (Fig.~\ref{jankahf3})
and thus much smaller than the stellar radius 
(Eq.~\ref{jankaheqrs}). Since the collapse proceeds very nearly
adiabatically ($T^3/\rho \approx {\mathrm{const}}$; 
Eq.~\ref{jankaheqs}), one can also easily estimate how the maximum
core temperature, which determines the neutrino emission, depends 
on the mass of the star. Mass conservation implies for
the final density $\rho_{\mathrm{f}} \sim 
(R_{\mathrm{crit}}/R_{\mathrm{s}})^3\rho_{\mathrm{crit}}$ with
$R_{\mathrm{crit}}/R_{\mathrm{s}}\propto M^{1/2}$ 
(Eq.~\ref{jankaheqrs}) and $\rho_{\mathrm{crit}}\propto 
M^{-7/2}$ (Eq.~\ref{jankaheqrhocrit}). Using this in 
Eq.~(\ref{jankaheqs}) one finds for the final temperature
$T_{\mathrm{f}} \propto M^{-1/2}$~\cite{Shi98}.
Neutrinos are emitted with a mean energy 
$\avejankah{\epsilon_{\nu}}$ that scales with the stellar
temperature in the main production region. A detailed 
discussion yields $\avejankah{\epsilon_{\nu}}\sim 
6kT_{\mathrm{f}}$~\cite{Shi98}.

The maximum neutrino plus antineutrino luminosity decreases 
steeply with higher stellar masses:
\begin{equation}
L_{\nu\bar\nu}\,\sim\, Q_{\nu}\,{4\pi\over 3}\,R_{\nu}^3
\propto \cases{M^{-3/2} &for $10^5\,M_{\odot} \lajankah 
M \lajankah 5\times 10^6\,M_{\odot}\,$,\cr
M^{-7} &for $5\times 10^7\,M_{\odot} \lajankah M \lajankah 
10^8\,M_{\odot}\,$.}
\end{equation}
The right plot in Fig.~\ref{jankahf3} shows this trend. The
change in the slope of the luminosities as a function of 
time that is visible for the cases $M = 10^7\,M_{\odot}$ and 
$M = 10^8\,M_{\odot}$ near 
$L_{\nu\bar\nu} \sim 10^{43}\,$ergs$\,$s$^{-1}$ is associated 
with the transition from the photo-neutrino dominated to the
pair-neutrino dominated regime (compare Fig.~\ref{jankahf2}).
The total energy release in neutrinos and antineutrinos 
exceeds $10^{56}\,$ergs for SMS with $M\lajankah 10^6\,M_{\odot}$.
Stars with smaller masses are the clearly stronger neutrino
sources:
\begin{equation}
E_{\nu}\,\sim\, L_{\nu\bar\nu}\, t_{\mathrm{em}}
\propto \cases{M^{-1/2} &for $10^5\,M_{\odot} \lajankah
M \lajankah 5\times 10^6\,M_{\odot}\,$,\cr
M^{-6} &for $5\times 10^7\,M_{\odot} \lajankah M \lajankah
10^8\,M_{\odot}\,$.}
\end{equation}
As displayed in Fig.~\ref{jankahf4}, the observable energies
are somewhat lowered by effects due to Doppler shift and 
gravitational redshift. SMSs near the lower end of the 
investigated mass range convert a fraction of more than
$10^{-4}$ of their rest-mass energy to neutrinos, whereas
it is less than $10^{-10}$ in case of stars with
$M = 10^8\,M_{\odot}$ (right plot in Fig.~\ref{jankahf4}).

\section{Conclusions}

Although the energy emitted in neutrinos is huge
in case of SMSs that form SMBHs with masses 
$M\lajankah 10^7\,M_{\odot}$,
it is very unlikely that these neutrinos can produce a 
highly relativistic outflow to power cosmic gamma-ray bursts.
On the one hand, the efficiency of neutrino-antineutrino
annihilation to electron-positron pairs is extremely low
($\lajankah 10^{-5}$ of the neutrino energy are converted
to $e^\pm$ pairs~\cite{Lin00,Lin01}). On the other hand,
99.8\% of this energy are deposited in the close vicinity
of the neutrino emitting shell, i.e. in matter which is 
swept inward in the collapse with velocities up to 60\%
of the speed of light. The deposited energy is much too 
small to invert this rapid infall and to create a hightly
relativistic outflow. A black hole-disk configuration 
might provide a more suitable environment, but is very
unlikely to form even in the case of the collapse of SMSs with 
rotation~\cite{Sai02}. Moreover, SMSs as gamma-ray
burst sources are unable to account for the short-time
variability of the observed emission, which is directly
linked to the activity of the source~\cite{Sar97}.
This requires a very compact energy source with a size 
which is typical of a neutron star or stellar mass black hole.

Investigations of the evolution and collapse of nonrotating 
SMSs are a somewhat academic exercise. Since the configurations
are so close to dynamical instability, a small amount of 
rotation may have a significant impact. This has indeed been 
found for the quasi-static evolution of uniformly 
rotating~\cite{Bau99a} and differentially
rotating stars~\cite{New01}. The collapse of 
stars which rotate uniformly at the onset of gravitational 
instability, however, turns out to be very similar to
the nonrotating case. Saijo et al.~\cite{Sai02} found that such a
collapse is likely to form a SMBH coherently, with almost all
of the matter falling into the hole on a dynamical timescale
and only very little matter possibly ending up in a disk.
They did not discover an unstable growth of a nonaxisymmetric
bar. Certainly such investigations of the death of SMSs 
should be extended to differentially rotating configurations and
to models which include a detailed microphysical description 
of the equation of state (instead of using a simple $\Gamma$-law
equation $P = (\Gamma-1)\varepsilon$). Moreover, the neutrino 
emission and possible energy release by nuclear burning
should be taken into account. The latter
might be more important for rotating SMSs than for nonrotating
ones. Centrifugal forces might hold up the collapse
long enough for nuclear reactions to generate sufficient 
energy to influence the dynamics even in case of zero 
initial metallicity~\cite{Ful86}.

\bigskip\noindent
{\bf Acknowledgements.} 
Stimulating conversations with T.~Abel and A.~Heger
are acknowledged. The author thanks F.~Linke, J.A.~Font, 
E.~M\"uller and P.~Papadopoulos for a pleasant 
collaboration and is very grateful to R.~Sunyaev for inspiring
discussions and 
the opportunity to present this review at the conference 
``Lighthouses of the Universe''. This work was supported by
the Sonderforschungsbereich 375 ``Astroparticle Physics" of the
Deutsche Forschungsgemeinschaft.

%

\end{document}